\documentclass[11pt]{article}

\usepackage{epsfig}

\def \sect #1 {\setcounter{equation} 0\section{#1}}
\def \be  {\begin{equation}}
\def \ee  {\end{equation}}
\def \ba  {\begin{eqnarray}}
\def \ea  {\end{eqnarray}}
\def \baa {\begin{eqnarray*}}
\def \eaa {\end{eqnarray*}}
\def \bb  {\begin{thebibliography}}
\def \eb  {\end{thebibliography}}
\def \as {\relax\ifmmode\alpha_s\else{$\alpha_s${ }}\fi}

\newcommand \bea{\begin{eqnarray}}
\newcommand \eea{\end{eqnarray}}

\def \O {\Omega}
\def \o {\omega}

\begin{document}

\begin{flushright}
YITP-SB-02-79 \\
\today
\end{flushright}

\vspace*{6mm}
\begin{center}
{\Large \bf Interjet Energy Flow/Event Shape Correlations}

\par\vspace*{7mm}\par

{\large
Carola F.\ Berger\footnote{Based on talk given by C. F. Berger at TH-2002,
 International Conference on Theoretical Physics,
Theme
2:``QCD, Hadron dynamics, etc.'', Paris, France, 2002.},
     Tibor K\'ucs, and
George Sterman}

\par\vspace*{5mm}\par

{\em C.N.\ Yang Institute for Theoretical Physics,
SUNY Stony Brook\\
Stony Brook, New York 11794 -- 3840, U.S.A.}

\vspace*{5mm}

\begin{abstract}
We identify a class of perturbatively computable measures of
interjet energy flow, which can be associated with well-defined
color flow at short distances. As an illustration, we calculate
correlations between event shapes and the flow of energy, $Q_\O$,
into an interjet angular region, $\O$, in high-energy two-jet
$e^+e^-$-annihilation events.  Laplace transforms with respect to
the event shapes suppress states with radiation at intermediate
energy scales, so that we may compute  systematically logarithms
of interjet energy flow.  This method provides a set of
predictions on energy radiated between jets, as a function of
event shape and of the choice of the region $\O$ in which the
energy is measured. Non-global logarithms appear as corrections.
We apply our method to a continuous class of event shapes.
\end{abstract}

\end{center}

\section{Introduction}

Inclusive jet cross sections have a long history, and the
agreement of predictions based on perturbative QCD  with
experiment is often impressive.  The study of energy flow into the
regions between energetic jets \cite{coflow} gives important, and
in some sense complementary information about the formation of
final states.  Interjet energy flow is expected  to reflect the
flow of color at short distances \cite{coflow,KOS, BKS1} and may
give insight into the hadronization process.  In addition,
knowledge of the relation between energy and color flows is important
for investigating hard collisions \cite{uevent}, to distinguish
QCD bremsstrahlung originating in the underlying hard scattering
or decay from radiation induced by multiple scatterings.  The
study of observables in jet events that test the flow of energy is
thus of considerable interest. The computation of such
observables, however, has turned out to be subtle \cite{DS}, for
reasons that we will briefly review below. For definiteness, we
will discuss these issues in dijet events at large center-of-mass
energy $Q$ in $e^+e^-$-annihilation \cite{BKS2}. In this case the
underlying color flow at short distances is unique, corresponding
to the creation of a quark-antiquark pair. This enables us to
illustrate our method, while avoiding  the complication of multiple
color exchanges, for example, singlet and octet exchange in
quark-antiquark scattering \cite{KOS,BKS1}.

\begin{figure}[htb]
%\vspace*{5cm}
\begin{center}
\epsfig{file=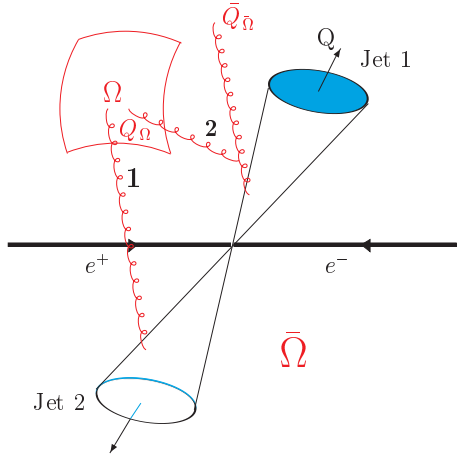,height=6.6cm,clip=0}
%\vbox{\vskip 1 in}
\caption{Sources of global and non-global logarithms in dijet
events. Configuration 1, a primary emission, is the source of
global logarithms, configuration 2 results in non-global
logarithms.} \label{eventfig}
\end{center}
\end{figure}

We study the flow of energy or transverse energy,
$Q_\O\equiv \varepsilon Q$, with $\varepsilon\ll 1$, into an
interjet angular region, $\O$. The process
\begin{equation}
e^+ + e^- \rightarrow \mbox{ 2 Jets }  + X_{\bar{\O}}
    + R_\O (Q_\O)\, ,
\label{event}
\end{equation}
is illustrated in Fig. \ref{eventfig}. $X_{\bar\O}$ represents soft
radiation into the region
between $\O$ and the jet axes, denoted by $\bar{\O}$, and $R_\O$ denotes
soft radiation into $\O$. Our goal is to relate soft
radiation into region $\O$ to the hard scattering that produces the jets.

Following
\cite{DS}, we refer to observables as ``non-global" when
they are defined by radiation into a specific portion of phase space, while
remaining more-or-less inclusive elsewhere.
In non-global cross sections there are two sources of logarithmic
corrections, as indicated in Fig.\ \ref{eventfig}: ``primary''
emissions, such as gluon 1 in Fig.\ \ref{eventfig}, are emitted
directly from the hard partons into $\O$. Phase space integrals
for these emissions contribute single logarithms per loop:
$\as^n\ln^n (Q/Q_\O) = \as^n \ln^n (1/\varepsilon)$.  These
logarithms exponentiate, and may be resummed in a straightforward
fashion \cite{BKS1}.  There are also ``secondary" emissions
originating from the complementary region $\bar{\O}$, illustrated
by configuration 2 in Fig.\ \ref{eventfig}. As emphasized by
Dasgupta and Salam \cite{DS}, emissions into $\O$ from such
secondary partons can give logarithms of the form $\as^n \ln^n
(\bar{Q}_{\bar{\O}}/Q_\O)$, where $\bar{Q}_{\bar{\O}}$ is the
maximum energy of radiation in $\bar{\O}$.  These have become known as
non-global logarithms. If no restriction is
placed on the radiation into $\bar{\O}$, then $\bar{Q}_{\bar{\O}}$
can approach $Q$, and the non-global, secondary logarithms can
become as important as the primary logarithms.  The non-global
logarithms arise because real and virtual enhancements associated
with secondary emissions do not cancel each other fully at fixed
$Q_\O$.  In general,  such non-global logarithms, as studied in
\cite{DS} and \cite{non-global}, are sensitive to color flow at
all scales and in all directions, and they do not exponentiate in
the same simple manner as do logarithmic enhancements from primary
emission.  In a sense, they mask the underlying color exchange at
short distances.

In the studies of non-global observables
in Refs.\ \cite{DS,non-global},  numerical estimates of both types of
logarithms were carried out in various situations.  Here we will adopt a
somewhat different
strategy, and introduce correlations between energy flow and event shapes
(``flow/shape correlations'').  Our aim is to select final  states
whose formation is sensitive primarily to
radiation from the highest-energy jets.
By restricting the range of
event shapes, we can systematically limit radiation in region $\bar{\O}$ while
retaining the chosen jet structure. This allows us to treat all leading
effects analytically.  In essence, the color and energy flow
originating from the primary radiation is less obstructed by
secondary radiation.
Our approach does not resum non-global logarithms, but allows us to
address the question
of interjet energy flow within a more global context, while
retaining the original motivation and physical picture described above.

\section{Energy Flow/Event Shape Correlations}

The following observation allows us to avoid leading contributions from
non-global configurations such as gluon 2 in Fig.\ \ref{eventfig}.
Large non-global logarithms originate from emissions into $\bar \O$ at
relatively large angles  from the jets, because radiation
from  partons close to
the jet directions is emitted coherently. By fixing the value of an event
shape near the
   two-jet limit, we avoid
final states with large energies in $\bar{\O}$ at fixed angles
from the jet axes. To impose this two-jet condition at fixed
energy flow, we will introduce shape functions $\bar{f}_c,\, c =
1,\,2$, one for each jet, for radiation into $\bar \O$. The
relevant cross section for the process (\ref{event}), at fixed
direction for jet 1, $\hat n_1$, is then defined as
\ba
{d \bar{\sigma}(\varepsilon,\bar{\varepsilon})\over d \varepsilon
\, d\bar{\varepsilon}\, d\hat n_1} &=& \frac{1}{2 Q^2}\ \sum_N\;
|M(N)|^2\, (2\pi)^4\, \delta^4(p_I-p_N)\;
\nonumber\\
&\ & \hspace{10mm} \times \ \delta(\varepsilon-f(N))\,
\delta(\bar{\varepsilon} -\bar{f}_1(N) - \bar f_2(N))\;
\delta(\hat n_1 -\hat n(N))\, ,
\label{eventdef}
\ea
with the
energy flow fixed by \ba f(N) & =  & \frac{1}{ Q}\ \sum_{\hat
n_i\in\O} \o_i\, .
\label{eflowdef}
\ea
We sum over all final
states $N$ that contribute to the weighted event, with $M(N)$ the
amplitude for ${\rm e^+e^-}\rightarrow N$. The total momentum is
$p_I$, with $p_I^2\equiv Q^2$. To define the total event shape
$\bar\varepsilon$ in Eq.\ (\ref{eventdef}) we divide $\bar \O$ into
two hemispheres, each surrounding one of the jets. Each hemisphere
contributes separately, through $\bar f_1$ and $\bar
f_2$. An example of a suitable shape function is the
thrust-like weight
\ba
\bar f_c(N) &=& \frac{1}{Q} \sum_{\hat
n_i\in \bar\O_c} \o_i \left( 1 - \cos \theta_i \right) , \,\, c =
1,2\, ,
\label{thrustex}
\ea
where $\theta_i$ are the angles with
respect to the jet directions. The cross section
Eq.\ (\ref{eventdef}) measures the
correlation of
$\bar \varepsilon = \bar{f}_1 + \bar{f}_2$
with the energy flow into $\O$. Since we are interested in two-jet
cross sections, we fix the constants $\varepsilon$ and
$\bar{\varepsilon}$ to be much less than unity:
\be
0 <
\varepsilon,\bar{\varepsilon} \ll 1.
\label{limit}
\ee
In this limit, we can neglect recoil due to the
soft radiation.
  Generalizations to
more than two jets in the final state, and to correlations with transverse
energy flow into $\O$, are clearly possible.

\begin{figure}[htb]
%\vspace*{5cm}
\begin{center}
a) \hspace*{5.4cm} b) \hspace*{5.4cm} c) \hspace*{4.5cm} \vspace*{-6mm}
\\
\epsfig{file=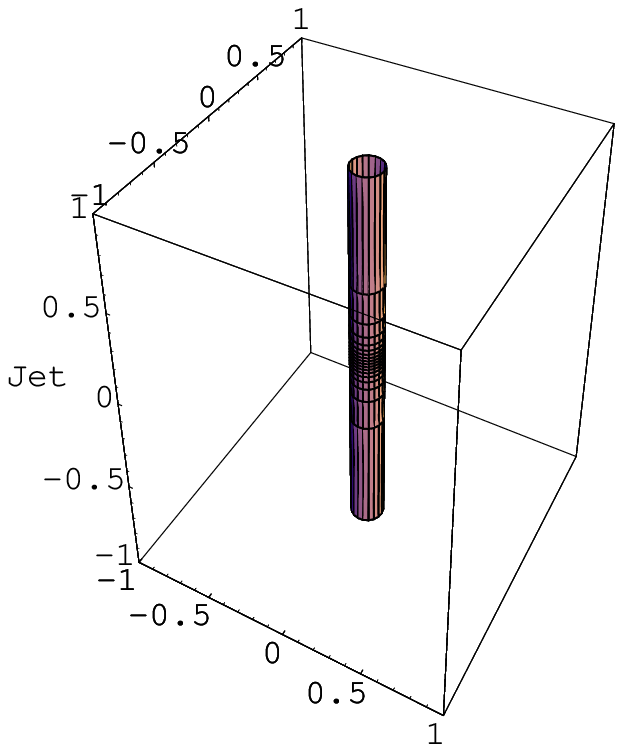,width=4cm,clip=0}  \mbox{ }\hspace*{15mm}
\epsfig{file=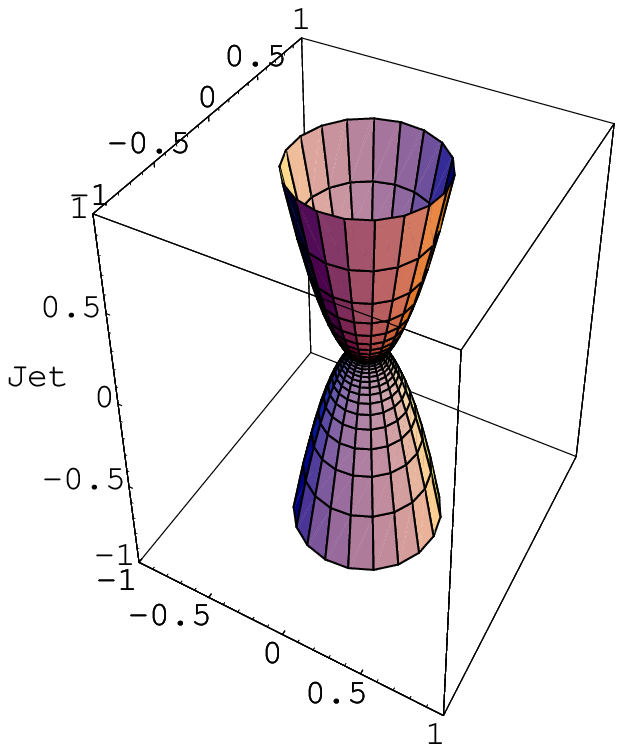,width=4cm,clip=0} \mbox{ }\hspace*{15mm}
\epsfig{file=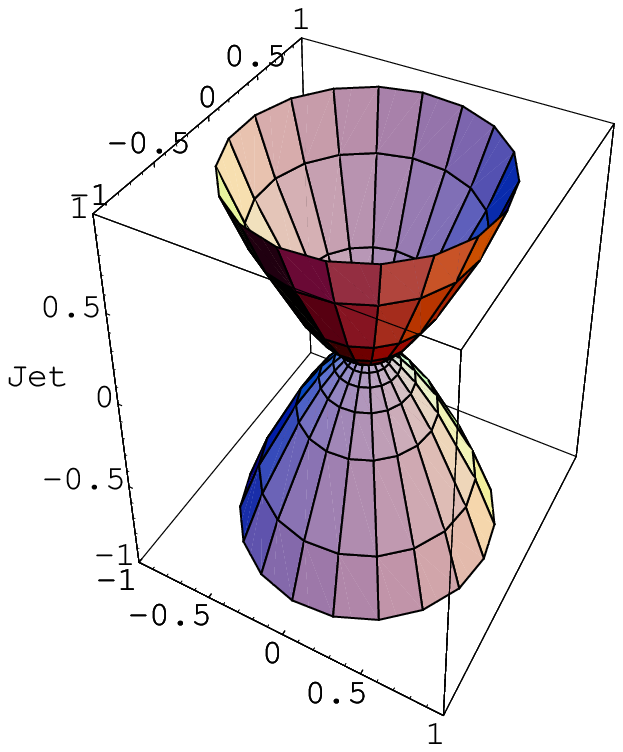,width=4cm,clip=0} %\vbox{\vskip 1 in}
\caption{Illustration of the effect of the parameter $a$ in the
weight (\ref{fbarexp}) on the shape of the event: a) shape for $a
= 1$, b) shape for $a = 0$, c) shape for $a = -1$. The jet axes
are in the vertical direction. The radial normalization
($\bar{\varepsilon} Q$) is arbitrary, but the same for all three
plots.} \label{shapefig}
\end{center}
\end{figure}

In the limit (\ref{limit}) the cross section (\ref{eventdef})
has large logarithmic enhancements in $\ln 1/\varepsilon$ and
$\ln 1/\bar{\varepsilon}$, which we will resum.
  However, we will
not resum logarithms like  $\ln(\bar{\varepsilon}/\varepsilon)$.
Instead, the following generalization of Eq. (\ref{thrustex}) to the
weights $\bar f_c(N,a)$ allows us to study
correlations of jet structure with energy flow into $\O$:
\ba \bar f_c(N, a) =
\frac{1}{Q}\ \sum_{\hat n_i\in \bar \O_c} \o_i \sin^a \theta_i
\left( 1 - \cos \theta_i  \right)^{1-a},\,\, c = 1,2, \quad a <
2. \label{fbarexp}
\ea
As $a \rightarrow 2$ the weight vanishes
only  very slowly for $\theta_i \rightarrow 0$, and at fixed $\bar
f_c$, the jet becomes very narrow. On the other hand, as $a
\rightarrow - \infty$, the event shape vanishes
in nearly every direction, and the cross section at fixed
$\bar f_c$ becomes more and more inclusive in radiation into
$\bar{\O}$. In this limit, non-global logarithms reemerge as
leading effects.  For $a = 0$ we obtain the thrustlike weight
(\ref{thrustex}) discussed above; the case $a=1$
corresponds to the jet broadening. The effect of $a$ on the shape
of the radiation into $\bar \O$ is illustrated in Fig.
\ref{shapefig}. In Fig. \ref{shapefig} we compare  the phase
space available to a particle at fixed $\bar \varepsilon$ for
three different values of $a$. The radial magnitude of each plot
is the maximum energy $\o$ found from Eq. (\ref{fbarexp}):
$r = \bar{\varepsilon} Q \sin^{-a} \theta (1-\cos \theta)^{a-1}$.
For $a = 1$, as shown in Fig. \ref{shapefig} a), the particle is
restricted to be close to the jet axes, while for $a = 0$,
and $a = -1$, depicted in Figs. \ref{shapefig} b) and c),
respectively, the particle is allowed to be farther away
from the axes.
For $a\ge 1$, recoil of the jets against
soft radiation cannot be neglected \cite{dok98}, so that
our analysis applies in the range $a<1$.
With the above choices of weight functions the cross section
(\ref{eventdef}) is infrared safe, and logarithms of
$\varepsilon$ and $\bar{\varepsilon}$ may be resummed.

\section{Factorization and Resummation}

\subsection{Factorization}

\begin{figure}[htb]
%\vspace*{5cm}
\begin{center}
\epsfig{file=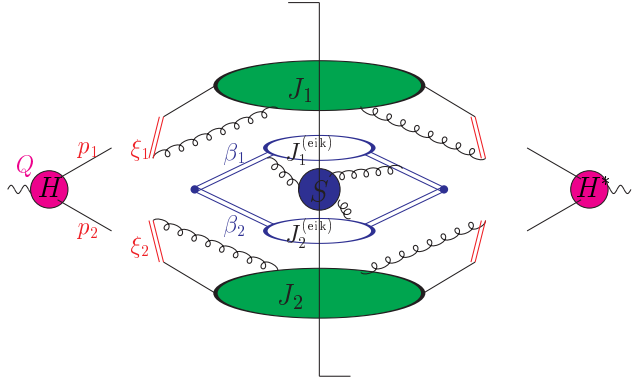,height=5cm,clip=0}
%\vbox{\vskip 1 in}
\caption{Factorized
cross section (\ref{factor}). The vertical line denotes the final
state, separating the amplitude (to the left) and
the complex conjugate amplitude (to the right).}
\label{factorized}
\end{center}
\end{figure}

Following the procedure developed and extensively discussed in
Refs.\ \cite{power,ColSte,ColSop,pQCD}, we first identify the
``leading regions", which can give rise to logarithmic
enhancements associated with lines approaching the mass shell.
When we sum over all diagrams that have a given final state, the
contributions from these leading regions may be factorized into a
set of functions, each of which corresponds to one of these
generic subdiagrams.  The cross section, Eq. (\ref{eventdef}),
then becomes a convolution in $\bar{\varepsilon}$, between
functions $\bar{J}_c$ that represent the evolution of the primary
partons, and a function $\bar{S}$ that represents coherent soft
radiation: \ba {d
\bar{\sigma}(\varepsilon,\bar{\varepsilon},a)\over d \varepsilon
\,d\bar{\varepsilon}\, d\hat n_1} &=& {d \sigma_0 \over
d\hat{n}_1}\ H(Q,\xi_1,\xi_2,\hat{n}_1,\mu)\; \int
d\bar{\varepsilon}_s\,
\bar{S}(\varepsilon,\bar{\varepsilon}_s,a,\mu) \,
\nonumber\\
&\ & \times \prod_{c=1}^2\, \int  d\bar{\varepsilon}_{J_c}\,
\bar{J}_c(\bar{\varepsilon}_{J_c},p_{J_c},\xi_c,a,\mu)\,
\delta(\bar{\varepsilon}-
\bar{\varepsilon}_{J_1}-\bar{\varepsilon}_{J_2}-
\bar{\varepsilon}_s)\, , \label{factor} \ea with $p_{J_c}$ the jet
momenta. $\mu$ is the factorization scale, which for simplicity we
set equal to the renormalization scale.
$d\sigma_0/d\hat{n}_1$ is the Born amplitude for the production of
a single particle (quark or antiquark) in direction $\hat{n}_1$.
The `short-distance' function
$H(Q,\xi_1,\xi_2,\hat{n}_1,\mu)=1+{\cal O}(\alpha_s(\mu^2))$,
which describes corrections to the hard scattering, is an
expansion in $\alpha_s$ with finite coefficients. In Eq. (\ref{factor}) we
have suppressed some  arguments
  of the soft function that are
not relevant for the discussion below.
The factorized
form (\ref{factor}) is illustrated in Fig. \ref{factorized}.

In (\ref{factor}),  the ``jet functions" $\bar{J}_c$ model
radiation collinear to the primary partons, with directions $+
\hat n_1$ for jet 1, and $-\hat n_1$ for jet 2, respectively,  and
are constructed to be independent of $\varepsilon$, which measures
the radiation at wide angles into $\O$.    The
vectors $\xi_c$ are introduced in
the factorization of the jet functions from the
hard scattering with the help of Ward identities.   For further
details we  refer to \cite{pQCD} and to \cite{BKS2}, where the jet
functions are related to QCD matrix elements, following
\cite{ColSop}.

The soft function $\bar{S}$ describes soft radiation at wide
angles from the jets, into $\O$ and $\bar \O$ alike, and thus
depends on both $\varepsilon$ and $\bar{\varepsilon}$. From the
arguments given in \cite{BKS1, pQCD}, it follows that radiation
at wide angles from the primary hard partons decouples from the
jets, and can be approximated by an eikonal cross section
$\sigma^{(\mbox{\tiny eik})}$, built out of path-ordered
exponentials $\Phi$:
\be
\Phi^{\rm (f)}_\beta
(\infty,0;x)
\equiv
   P e^{-i g \int_{0}^{\infty} d \lambda \beta
\cdot {\mathcal{A}}^{\rm (f)} (\lambda \beta + x )},
\ee
where the $\beta$'s are light-like velocities in the directions of the
jets. The superscript $\rm (f)$ indicates that the vector potential
takes values in representation f, in our case the
representation of a quark or an antiquark.  An eikonal cross
section  also
contains enhancements for configurations collinear to the jets,
which in the factorization (\ref{factor}) are  already taken
into account in the partonic jet functions. Therefore,
the eikonal cross section itself is not a suitable soft function.
Rather, to avoid
overcounting and to include only soft, but not collinear
enhancements, the soft function is defined through
the refactorization \cite{KOS}
\ba
\bar{\sigma}^{(\mbox{\tiny
eik})}\left(\varepsilon,\bar{\varepsilon}_{\mbox{\tiny eik}},a,\mu
\right) & \equiv  &
\int  d \bar{\varepsilon}_s \
\bar{S}\left(\varepsilon,\bar{\varepsilon}_s,a,\mu
\right)
\nonumber
\\
&\ & \times\
\prod\limits_{c = 1}^2 \; \int d\bar{\varepsilon}_c\;
\bar{J}_c^{(\mbox{\tiny
eik})}\left(\bar{\varepsilon}_{c},a,\mu\right)
    \delta \left(\bar{\varepsilon}_{\mbox{\tiny eik}} -
\bar{\varepsilon}_s - \bar{\varepsilon}_1 - \bar{\varepsilon}_2
\right). \label{eikfact}
\ea
In Eq.\ (\ref{eikfact}) we have
refactorized the eikonal cross section in the same manner as the
partonic cross section (\ref{factor}), into
eikonal analogs of jet functions,
and the same soft function \cite{BKS2}.
We have omitted the dependence of the functions in (\ref{eikfact})
on the eikonal lines $\beta_c$ and $\xi_c$ for better readability.

\subsection{Resummation}

To disentangle the convolution in (\ref{factor}), we take Laplace
moments with respect to $\bar{\varepsilon}$,
\ba
     {d \sigma(\varepsilon,\nu,a)\over d \varepsilon \,d\hat n_1}
& = &  \int_0^\infty d\bar{\varepsilon}\; e^{- \nu
\,\bar{\varepsilon}}\ {d
\bar{\sigma}(\varepsilon,\bar{\varepsilon},a)\over d \varepsilon\,
d\bar{\varepsilon}\, d\hat n_1} \label{trafo} \nonumber
\\
& = &  {d \sigma_0 \over d\hat{n}_1}\ H(\xi_1,\xi_2,\hat{n}_1,Q)\;
\prod_{c = 1}^2\; J_c\left(\frac{p_{J_c} \cdot \xi_c}{Q
\left|\xi_c\right|},\nu,a,\as(Q) \right)
    \  S\left(\varepsilon, \nu, a, \as(Q) \right)\, ,
    \label{trafosig}
\ea
where we have exhibited the relevant $\xi_c$-dependence
of $J_c$ \cite{BKS2}.
Here and below  unbarred quantities are the transforms in
$\bar{\varepsilon}$, and barred quantities are untransformed.
In Eq. (\ref{trafosig}) we have set the factorization scale to $Q$,
which avoids large logarithms
in the hard function. Large logarithms of $\varepsilon$ and $\nu$ or
$\bar{\varepsilon}$ then occur in the soft and jet functions. In addition, the
jets contain potentially large logarithms of ${p_{J_c} \cdot
\xi_c}/({\mu \left|\xi_c\right|})$.

The resummation of these logarithms follows
from the independence of the physical correlations from the
factorization scale and the choice of eikonal factors $\xi_c$,
\ba
\mu \frac{d}{d \mu} \frac{d \sigma \left(\varepsilon,\nu,a
\right)}{d\varepsilon\,d \hat{n}_1 } & = & 0, \label{muev}
\\
\frac{\partial}{\partial \ln  \left(p_{J_c} \cdot \xi_c\right) }
\frac{d \sigma \left(\varepsilon,\nu,a \right)}{d\varepsilon\,d
\hat{n}_1 } & = & 0. \label{xiev} \ea Following the methods
described in Refs.\ \cite{ColSop,pQCD,cls}, we can use
(\ref{muev}) and (\ref{xiev}) to derive evolution equations for
the jet and soft functions.  From this procedure, we derive a
resummed expression for the Laplace-transformed correlations \cite{BKS2},
which may be written as
  \ba
\frac{d \sigma \left(\varepsilon, \nu,a \right)}{d \varepsilon
\, d \hat{n}_1
} &=&
   {d \sigma_0 \over d\hat{n}_1}\
H(\xi_1,\xi_2,\hat{n}_1,Q) \,  S\left(\varepsilon \nu, a,\O,
\as(\varepsilon Q ) \right)  \, \exp \left[  -
\int\limits_{\varepsilon Q}^{Q/2} \frac{d
\lambda}{\lambda} \gamma_s\left(\as(\lambda)\right) \right]
\nonumber\\
&\ & \times
\prod_{c=1}^2\, J_c
\left(1,1,a,\as\left(\frac{Q}{2\,\nu^{1/(2-a)}}\right) \right)
    \exp \left\{ -  \int\limits_{\frac{Q}{2\,\nu^{1/(2-a)}}}^{Q/2}
\frac{d\lambda}{\lambda}  \gamma_{J_c}\left(\as(\lambda)\right) \right\}
\nonumber \\
& \ &\quad \times\
\exp \left\{- \int\limits_{\frac{Q}{2\,\nu^{1/(2-a)}}}^{Q/2}
  \frac{d \lambda}{\lambda}  \left[
B_c\left(a, \as(\lambda)\right) + \int\limits_{C_1\frac{Q^{2-a}}{\nu
(2 \lambda)^{1-a}}}^{C_2\lambda} \frac{d \lambda'}{\lambda'}
A_c\left(
a,\as(\lambda')\right) \right] \right\}\, ,
    \label{evolend}
\ea
where we have exhibited the relevant arguments of the soft function.
Convenient choices for the scales in the
$\lambda'$ integral are $C_2=1$ and $C_1=\exp[-\gamma_E+(1-a)/2]$.
The functions $A_c$ and $B_c$ are associated with
the jet functions \cite{ColSop,cls}, and the
anomalous dimensions $\gamma_s$ and $\gamma_{J_c}$
belong to the soft and jet functions, respectively
\cite{BKS2, ColSop}.  At lowest order, $A_c=2C_c(\alpha_s/\pi)$,
with $C_c=C_F$ for quarks and $C_A$ for gluons.  Other explict
forms are given in \cite{BKS2}.
Only the soft function retains dependence on the geometry of the
region $\O$, as explicitly indicated in (\ref{evolend}).
Corrections in Eq. (\ref{evolend}) of
the form $\alpha_s (Q) \ln 1/(\varepsilon \nu)$
occur entirely within the soft function $S$.   For large enough
$\nu$ and/or $|a|$, at fixed $\varepsilon$, large non-global
logarithms reemerge in that function, which may
be computed in the eikonal approximation.

So long as we are able to compute corrections in $\alpha_s(Q) \ln
1/(\varepsilon \nu)$ perturbatively,
   we are free to choose between $\varepsilon Q$ and $Q/\nu$ in
any term in (\ref{evolend}) that generates single logarithms. Of
particular relevance is the $\gamma_s$ integral in the first
exponential of (\ref{evolend}), which reflects the choice of
renormalization scale in the soft function $S$.  At higher orders
in $S$, logarithms will arise from gluon emission into  angular
regions for which the upper limit on energy flow does not match
the renormalization scale.  Within $\O$ the upper limit is
$\varepsilon Q$, and outside, in $\bar{\O}$, it is $Q/\nu$.
(Recall, there are no collinear enhancements in $S$.) These
logarithms appear multiplied by the size of the angular region
from which they arise.  When we measure energy flow into a region
$\O$ of small angular extent, it is more convenient to choose a
renormalization scale $Q/\nu$, to avoid generating
logarithms of $\varepsilon\nu$ multiplied by the size of the larger
region $\bar{\O}$. Correspondingly, when $\O$ subtends the bulk of
the unit sphere, it is convenient to make the choice of
$\varepsilon Q$, as in Eq.\ (\ref{evolend}).

\section{Results}

\begin{figure}[hb]
\vspace*{7mm}
\begin{center}
a) \hspace*{7.5cm} b) \hspace*{4cm} \vspace*{-6mm} \\
\epsfig{file=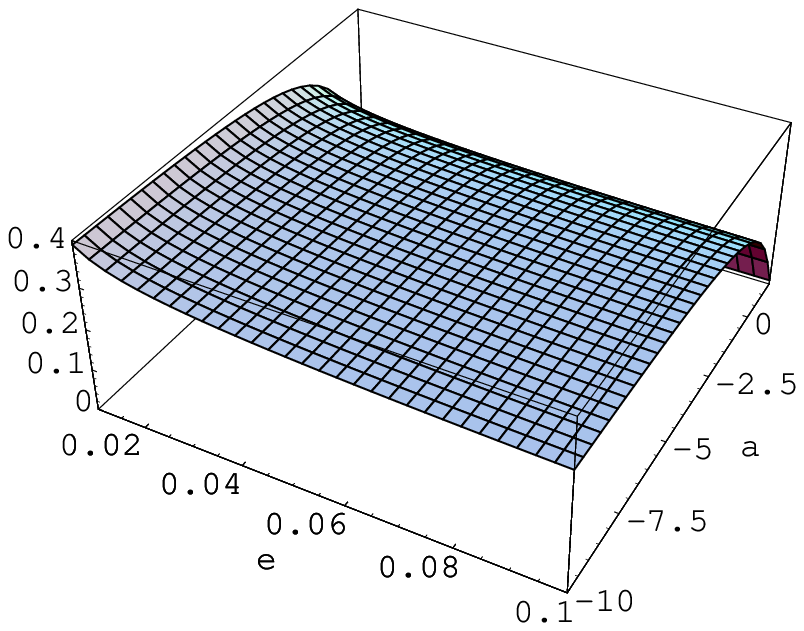,width=6.8cm,clip=0}  \mbox{
}\hspace*{5mm}  \epsfig{file=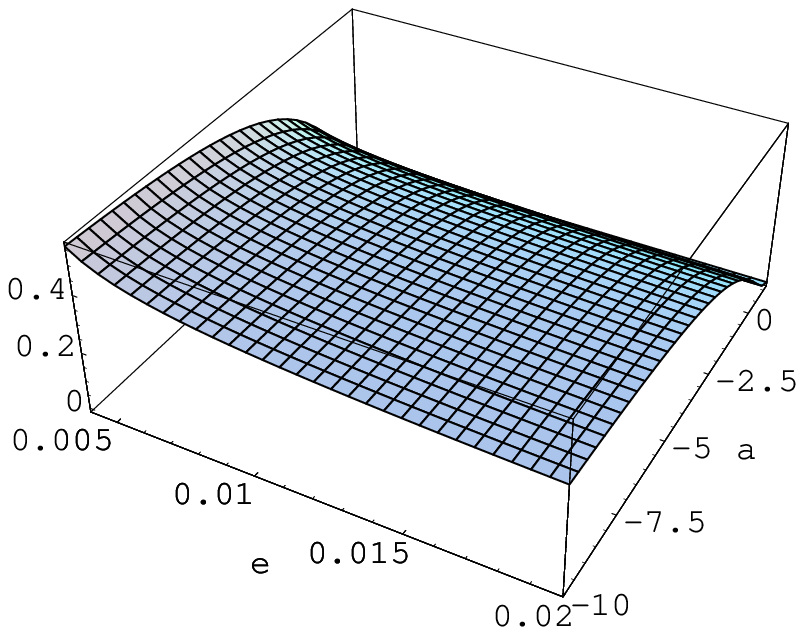,width=6.8cm,clip=0}
\vbox{\vskip 1 in}
\caption{Differential cross section $\frac{\varepsilon d
\sigma/(d\varepsilon d\hat n_1)}{d \sigma_0/d \hat n_1}$,
normalized by the lowest order cross section, at $Q = 100$ GeV,
as a function of
$\varepsilon$ and $a$ at fixed $\nu$: a) $\nu = 10$, b) $\nu =
50$. $\O$ is a slice (ring) centered around the jets, with a
width of $\Delta \eta = 2$.} \label{num1}
\end{center}
\end{figure}

\begin{figure}[htb]
\vspace*{6mm}
\begin{center}
a) \hspace*{7.5cm} b) \hspace*{4cm} \vspace*{-6mm} \\
\epsfig{file=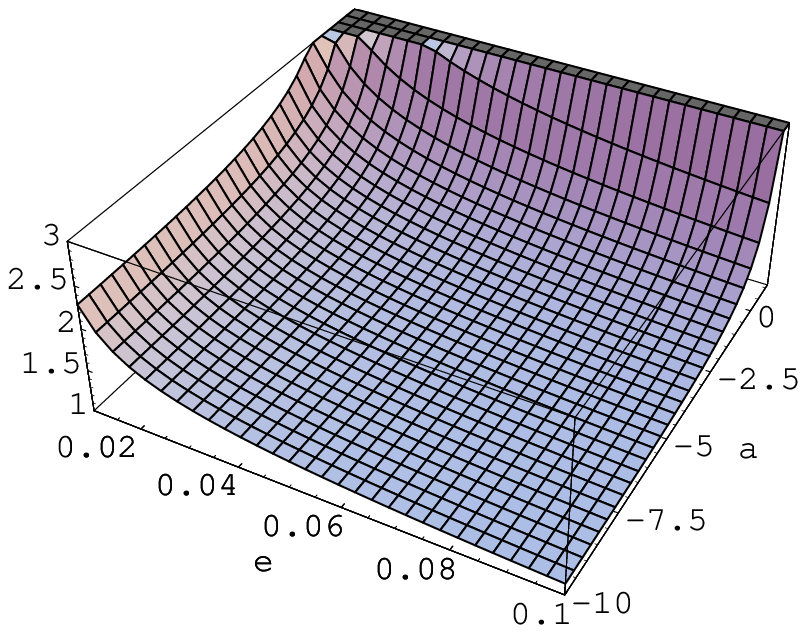,width=6.7cm,clip=0}  \mbox{
%\vbox{\vskip 1 in}
}\hspace*{5mm}  \epsfig{file=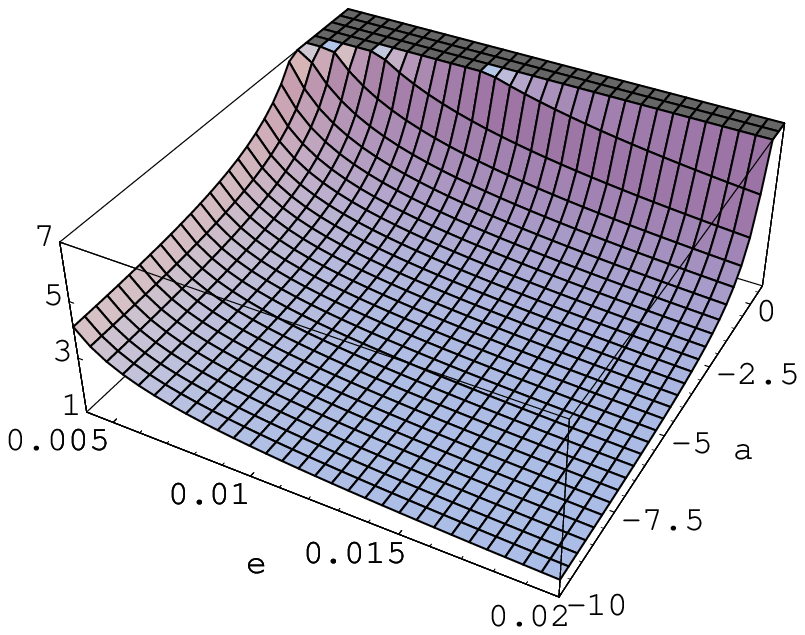,width=6.7cm,clip=0}
\caption{Ratios of differential cross sections  for quark to
gluon jets $\frac{C_A}{C_F} \left(\frac{\varepsilon d
\sigma^q/(d\varepsilon d\hat n_1)}{d \sigma_0^q/d \hat n_1}\right)
\left(\frac{\varepsilon d
\sigma^g/(d\varepsilon d\hat n_1)}{d \sigma_0^g/d \hat n_1}\right)^{-1}$ at $Q = 100$ GeV
as a
function of $\varepsilon$ and $a$ at fixed $\nu$: a) $\nu = 10$,
b) $\nu = 50$. $\O$, as in Fig. \ref{num1}, is a slice (ring)
centered around the jets, with a width of $\Delta \eta = 2$.}
\label{num2}
\end{center}
\end{figure}

Eq.\ (\ref{evolend}) resums single logarithms
of $\varepsilon$ and single and double logarithms of $\nu$. The
latter stem from the double integral in the last exponent
on the right-hand side. In Figs. \ref{num1} and \ref{num2} we show some
examples of numerical results from (\ref{evolend}).

Fig. \ref{num1} shows the dependence of the differential cross
section (\ref{evolend}), multiplied by $\varepsilon$
and normalized by the Born cross section,
$\frac{\varepsilon d \sigma/(d\varepsilon d\hat n_1)}{d
\sigma_0/d \hat n_1}$, on the measured energy $\varepsilon$ and
on the parameter $a$, at fixed $\nu$.  We choose the
region $\O$ to be a ring around the jets, centered in their
center-of-mass, with a width of $\Delta \eta = 2$. The center-of-mass energy $Q$
is chosen to be $100$ GeV.
In Fig. \ref{num1} a), we plot $\frac{\varepsilon
d \sigma/(d\varepsilon d\hat n_1)}{d \sigma_0/d \hat n_1}$ for
$\nu = 10$, in Fig. \ref{num1} b) for $\nu  = 50$.
The cross section falls as $a$ increases,
where the jets are restricted to
be very narrow. Similarly, as $\nu$ increases, the radiation into the
complementary region $\bar \O$ is more restricted,
as illustrated by the comparison of Figs.
\ref{num1} a)  and b).  On the other hand,
as $|a|$ increases at
fixed $\varepsilon$, the correlations (\ref{evolend}) approach a
constant value.  For $a\rightarrow -\infty$, however, non-global
dependence on
$\varepsilon$ and $|a|$ will emerge from higher order corrections
in the soft function.

To illustrate the sensitivity of these results to the flavor of
the primary partons, we study the corresponding ratio of the
correlation to the cross section for gluon jets produced
by a hypothetical color singlet source. Fig. \ref{num2} displays the ratio
of the differential cross section
$d\sigma^q(\varepsilon,a)/(d\varepsilon d \hat n_1)$, normalized
by the lowest order cross section, to the
analogous quantity with gluons as
primary partons in the outgoing jets, again at
$Q = 100$ GeV. We multiply the ratio by $C_A/C_F$ to compensate for
the difference in the normalization of the lowest order soft functions.
Gluon jets
are wider, and hence are
suppressed relative to quark jets as $a$ or $\nu$ increases. Figs. \ref{num2} a) and b)
exhibit this behavior, where we compute the cross
section for the same ring around the jets, centered in their
center-of-mass, with a width of $\Delta \eta = 2$.  Again, Fig.
\ref{num2} a) is at $\nu = 10$ and b) is at
$\nu  = 50$. These results suggest sensitivity to
the more complex color and flavor flow
characteristic of hadronic scattering.

%\clearpage

\section{Conclusions}

We have described a study of interjet energy flow/event shape
correlations in $e^+e^-$ dijet events, where all leading effects
are treated analytically. By weighting the final state radiation
appropriately, our formalism is sensitive mainly to radiation
stemming directly from the primary hard scattering. Transforms in
a class of weight functions enable us to control the influence of
secondary, or non-global, radiation on the energy flow between
the jets, and in principle to study correlations between jet
structure and energy flow.  The study of next-to-leading
order and higher corrections in which these effects
appear will also be of interest.   The application of our formalism
to multijet events and to scattering with initial state hadrons is
certainly possible, and should shed light on the relationship
between color and energy flow in hard scattering processes.

\subsection*{Acknowledgements}

We thank Gavin Salam for useful discussions.
C.F.B.\ thanks the National Science Foundation and the sponsors of
TH2002 for support at TH2002.  This work was also supported in part by
the National
Science Foundation grant PHY0098527.

\end{document}